\documentclass[12pt]{article}
\usepackage{epsfig}

\newcommand{\be}{\begin{equation}}
\newcommand{\ee}{\end{equation}}
\newcommand{\bq}{\begin{eqnarray}}
\newcommand{\eq}{\end{eqnarray}}
\newcommand{\one}{1 \!\! 1}
\newcommand{\D}{\mathrm{d}}
\newcommand{\E}{\mathrm{e}}
\newcommand{\I}{\mathrm{i}}

\def\slash{\rlap{/}}
\def\lsim{\mathrel{\rlap{\lower4pt\hbox{\hskip1pt$\sim$}}\raise1pt\hbox{$<$}}}
\def\gsim{\mathrel{\rlap{\lower4pt\hbox{\hskip1pt$\sim$}}\raise1pt\hbox{$>$}}}
\def\Vec#1{\mathpalette{\VVec}{#1}}                  
\def\VVec#1#2{\mbox{\boldmath$#1#2$\unboldmath}}
\def\nostrocostruttino#1\over#2{\mathrel{\mathop{\kern 0pt \rlap
{\hbox{$#1$}}} \hbox{\kern-.135em $#2$}}}

\newcommand{\Strut}{\rule[-1.7ex]{0pt}{4.7ex}}        
\newcommand{\STRUT}{\rule[-2.1ex]{0pt}{5.5ex}}        
\def\anti#1{\mathpalette{\@anti}{#1}#1}
\def\@anti#1#2{\sbox0{$#1#2$}
  \makebox[0pt][l]{$#1\kern.30\ht0\overline{\kern-.35\ht0\phantom{#2}}$}}

\begin{document}

\pagestyle{empty}

\null
\vspace{2cm}

\begin{center}
{\bf \Large SIDIS in the target fragmentation region:}
\vskip 6pt
{\bf \Large polarized and transverse momentum dependent}
\vskip 6pt
{\bf \Large fracture functions}

\end{center}

\vspace{1cm}

\begin{center}

{\large M.~Anselmino$^{a}$, V.~Barone$^b$, A.~Kotzinian$^{a,c}$}

\vspace{0.2cm}

$^a${\it Dipartimento di Fisica Teorica, Universit{\`a}
di Torino; \\
INFN, Sezione di Torino, 10125 Torino, Italy}

$^b${\it Di.S.T.A., Universit{\`a} del Piemonte
Orientale ``A. Avogadro''; \\
INFN, Gruppo Collegato di Alessandria,  15121 Alessandria, Italy}

$^c${\it Yerevan Physics Institute, 375036 Yerevan, Armenia}
\end{center}

\vspace{2cm}

\begin{center}

{\large ABSTRACT}

\end{center}

\vspace{0.5cm}

\noindent The target fragmentation region of semi-inclusive deep inelastic
scattering is described at leading twist, taking beam and target
polarizations into account. The formalism of polarized and
transverse-momentum dependent fracture functions is developed and the
observables for some specific processes are presented.

\vspace{0.5cm}


\newpage

\pagestyle{plain}

\section{Introduction}

The study of the 3-dimensional partonic structure of nucleons has become
in recent years a central issue in hadron physics, with impressive
dedicated theoretical and experimental activities. The ultimate goal
is that of achieving a 3-dimensional imaging of the nucleons, both in
configuration and momentum space.
The information on the momentum distributions of quarks and gluons
 is encoded in the Transverse Momentum Dependent
distribution functions (TMDs), which are probed  in inclusive
processes, mainly in Semi Inclusive Deep Inelastic Scattering (SIDIS,
$l \, N \to l \, h \, X)$.

In SIDIS TMDs can be accessed (at leading order in $\alpha_{QED}$)
studying the azimuthal modulations of the cross section around the
virtual photon direction (for a recent review see Ref.~\cite{Barone:2010zz}).
The detected final hadron is generated in the fragmentation of a
scattered quark, the so-called Current
Fragmentation Region (CFR). The cross section is then factorized as a
convolution of transverse momentum dependent distribution and
fragmentation functions; thus, in order to explore the parton momentum
structure of the nucleon one has to be sure to select events in the CFR.

However, final hadrons in SIDIS and other partially inclusive processes
can also be found among the remnants of the struck target, the so-called
Target Fragmentation Region (TFR). The appropriate QCD formalism developed
to study particle production in the TFR is that of the Fracture Functions,
introduced by Trentadue and Veneziano~\cite{Trentadue:1993ka} to describe
the partonic structure of a nucleon when it fragments into a final-state
hadron.

Although, in principle, the two regions can be
kinematically separated, some situations may not be so clear;
in particular at intermediate energies, when the total number of final
produced particles is not very high. It is then worth studying the
properties and the azimuthal distribution of hadrons produced in the TFR.
This could both serve as a test of our complete understanding of the
different mechanisms in the SIDIS production of hadrons and add
additional information on the features of hadrons produced in the
two regions. In some cases the same azimuthal dependences can be found
in the CFR and in the TFR; then, great care must be taken when analyzing
the data. Some other azimuthal dependences, instead, are typical and
unique for one of the two regions, strengthening their interpretation.

Let us briefly recall how the production mechanisms in the two regions,
CFR and TFR, can be described in semi-inclusive deep inelastic scattering
(SIDIS), $l(\ell) + N(P) \to l(\ell') + h(P_h) + X(P_X)$. Its kinematics
is described by the usual invariants ($q^{\mu}$ is the momentum of the
exchanged virtual photon)
\begin{equation}
  x_B = \frac{Q^2}{2P{\cdot}q} \quad \quad
  y = \frac{P{\cdot}q}{P{\cdot}\ell} \quad \quad
  z_h = \frac{P{\cdot}P_h}{P{\cdot}q} \quad \quad
  W^2 = (P + q)^2 \, \cdot
\label{variables}
\end{equation}
In the center of mass frame of the virtual photon and the nucleon (the
c.m. $\gamma^* N$ frame), with the photon directed along the positive $z$
direction, the variable $z_h$ is
\be
z_h \simeq \frac{P_h^+}{q^+}\,,
\label{zratio}
\ee
where the light-cone components of a generic vector $A^{\mu}$
are defined as $A^{\pm} \equiv (A^0 \pm A^3)/\sqrt{2}$.
In the ratio (\ref{zratio}) $q^+$ is always large, $q^+ \sim Q$, whereas
the magnitude of $P_h^+$ determines the fragmentation region:

\begin{tabular}{ll}
{\it Current fragmentation region} (CFR): & $P_h^+ \sim Q$\,, \\
{\it Target fragmentation region} (TFR): & $P_h^+ \sim 0$\,.
\end{tabular}
\vskip 12pt

Equivalently, defining in the c.m. $\gamma^* N$ frame the
hadron momentum as
$P_h = (E_h, \Vec P_{h\perp}, P_{h \parallel})$, the usual Feynman variable
$x_F = 2P_{h \parallel}/W$ identifies the CFR and the TFR, respectively, by
$x_F > 0$ and $x_F < 0$. A detailed discussion of the operational
criteria to separate the two regions can be found in Ref. \cite{Berger:1987}.

\begin{figure}[t]
\begin{center}
\includegraphics[width=0.40\textwidth]
{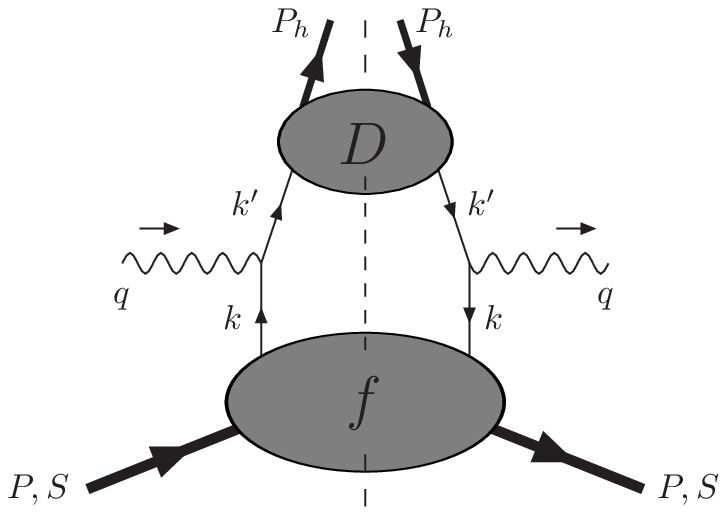}
\hspace{0.2cm}
\includegraphics[width=0.40\textwidth]
{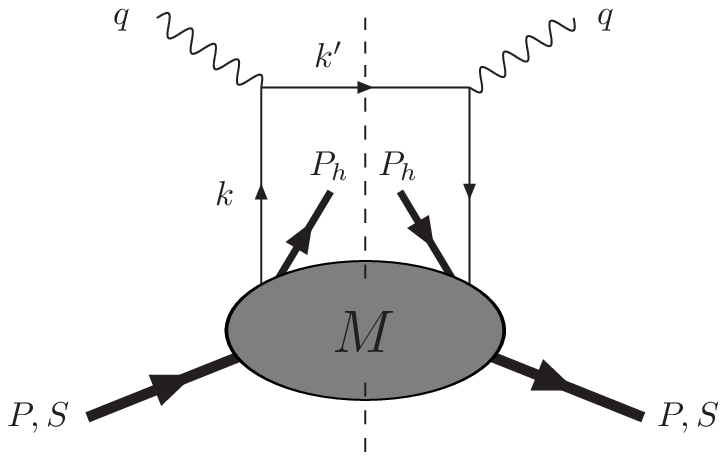}
\caption{The handbag diagram for the SIDIS hadronic tensor in the current
fragmentation region (left) and in the target fragmentation region
(right).}
\label{handbag}
\end{center}
\end{figure}

In the CFR the SIDIS cross section integrated over the transverse
momentum of the final hadron can be factorized at lowest order as
\be
\frac{\D \sigma^{\rm CFR}}{\D x_B \, \D y \, \D z_h}
= \sum_a e_a^2 \, f_a (x_B) \, \frac{\D \hat\sigma}{\D y} \, D_a (z_h) \,,
\label{sidis}
\ee
where $f_a(x_B)$ is the distribution function of parton $a$, $D_a(z_h)$
is the fragmentation function of parton $a$ into hadron $h$ and
$\D \hat{\sigma}/\D y$ is the elementary cross section of lepton-quark
scattering. The parton-model graph describing this process is the handbag
diagram shown in Fig.~\ref{handbag} (left). The partonic meaning of the two variables is the following: $x_B$ is the fraction of the longitudinal
momentum of the nucleon carried by the quark, $z_h$ is the fraction of the
longitudinal momentum of the struck quark carried by the final hadron (we
have dropped all scale dependences in the distribution and fragmentation
functions).

In the TFR the factorization in $x_B$ and $z_h$ of Eq.~(\ref{sidis})
does not hold any longer, as it is not possible to separate
the quark emission from the hadron production. Moreover,
$z_h$ is not the proper variable to describe this region. The reason
is easily understood if we write $z_h$ in the c.m. $\gamma^* N$ frame (we
neglect as usual hadron masses):
\be
z_h = \frac{E_h}{E (1 - x_B)}  \frac{(1 - \cos \theta_h)}{2} \,,
\label{zh}
\ee
where $\theta_h$ is the angle between $\Vec P_h$ and $\Vec P$.
The $z_h$ variable does not discriminate between two different
physical situations, namely $E_h = 0$ (soft hadron emission) and
$\theta_h = 0$ (target fragmentation: emission of a hadron collinear
with the target remnant), which both correspond to $z_h = 0$.

In order to  describe the production of hadrons in the target
fragmentation region, one has to define  the fracture functions
$M_a (x_B, (1 - x_B) z)$, which depend on $x_B$ and on a
new variable $z = E_h/E (1 - x_B)$, and represent the distributions
of partons inside a nucleon fragmenting almost collinearly into
a given hadron \cite{Graudenz:1994dq,deFlorian:1997wi}.
Notice that, differently from $z_h$, the variable $z$
vanishes in the soft limit only ($E_h \to 0$).
The SIDIS cross section in the TFR, integrated over the transverse
momentum of the final hadron, thus becomes

\be
\frac{\D \sigma^{\rm TFR}}{\D x_B \, \D y \, \D z}
= \sum_a e_a^2 \, (1 - x_B)\,  M_a (x_B, (1 - x_B) z)
\, \frac{\D \hat\sigma}{\D y} \,\cdot
\label{sidis2}
\ee

In the following we will consider SIDIS processes in the TFR and
will develop the formalism of fracture functions for polarized SIDIS
and for $\Vec P_{h \perp}$ distributions. We will introduce and
classify all the leading-twist transverse momentum dependent and
polarized fracture functions and present the lowest-order results
for cross sections and angular distributions of single-hadron
lepto-production with polarized beam and/or target.

\section{SIDIS in the target fragmentation region}

The SIDIS cross section is, in the one photon exchange approximation,
given by the contraction of a leptonic tensor $L^{\mu \nu}$
with the hadronic tensor $W^{\mu \nu}$ incorporating the
structure of the target nucleon and the dynamics of fragmentation:
\begin{equation}
  2 E_h \, \frac{\D \sigma}{\D x_B \, \D y \,  \D^3\Vec{P}_h}
  =
  \frac{\pi \, \alpha_{\rm em}^2 \, y}{Q^4} \, L_{\mu\nu} W^{\mu\nu} \, .
  \label{sidis8}
\end{equation}

It is convenient to use a Sudakov parametrization  of the
relevant momenta. We introduce two null vectors, $p^{\mu}$ and $n^{\mu}$,
with $p \cdot n = 1$, $p^- = P^-$, $n^+ = 1/P^-$, $p^+ = n^- = 0$,
and we work in a $\gamma^* N$ frame. The unit vector $\hat{\Vec q}
\equiv \Vec q /\vert \Vec q \vert$ identifies the positive
$z$ direction. In terms of the Sudakov vectors $p^{\mu}$ and $n^{\mu}$
the  four-momenta at hand are:
\bq
& &  P^{\mu} = p^{\mu} + \frac{m_N^2}{2} \, n^{\mu}
\simeq p^{\mu}  \\
& &  q^{\mu} \simeq \frac{Q^2}{2 x_B} \, n^{\mu} - x_B \, p^{\mu} \\
& & P_h^{\mu} = \zeta \, p^{\mu} + \frac{\Vec P_{h \perp}^2 + m_h^2}{2 \zeta}
\, n^{\mu}
 + P_{h \perp}^{\mu} \simeq \zeta \, p^{\mu} + P_{h \perp}^{\mu}
 \,.
\eq
The momentum of the hadron is identified by
the light-cone ratio $\zeta = P_h^-/P^- \simeq E_h/E =
z (1 - x_B)$ and its transverse component $P_{h \perp}^{\mu}
= (0, \Vec P_{h \perp}, 0)$, with an azimuthal angle $\phi_h$ in
the plane perpendicular to the $\gamma^* N$ axis. Note that in the
high-energy limit one has $\zeta \simeq P_{h \parallel}/P_{\parallel}
\equiv x_L
\simeq (1 - x_B) \vert x_F \vert$.
Replacing $\Vec P_h$ with the variables $(\zeta, \Vec P_{h \perp})$
 and allowing for  target polarization (see Ref.~\cite{Diehl:2005pc}),
the cross section takes the form
\be
  \frac{\D \sigma}{\D x_B \, \D y \, \D \zeta \,
\, \D^2 \Vec P_{h \perp} \, \D \phi_S}
  =
  \frac{\alpha_{\rm em}^2}{4 \, Q^4} \, \frac{y}{\zeta} \,
L_{\mu\nu} W^{\mu\nu} \,.
  \label{sidis9}
\ee
Here $\phi_S$ is the azimuthal angle of the transverse component
of $S^{\mu}$, the nucleon spin vector, parametrized as
\be
S^{\mu} =
S_{\parallel} \, \frac{p^{\mu}}{m_N} - S_{\parallel}
\, \frac{ m_N}{2} \,
n^{\mu} + S_{\perp}^{\mu}
\simeq S_{\parallel}\, \frac{p^{\mu}}{m_N} + S_{\perp}^{\mu}\,.
\ee
A similar decomposition holds for
the spin vector of the produced hadron $h$:
\be
S_h^{\mu}
= S_{h \parallel} \frac{ \zeta \, p^{\mu}}{m_h}
- S_{h \parallel} \frac{ m_h}{2 \zeta} \, n^{\mu} + S_{h \perp}^\mu
\simeq S_{\parallel} \, \frac{ \zeta \, p^{\mu}}{m_h}  +
S_{h \perp}^{\mu} \,.
\ee
Strictly speaking one should distinguish between transverse vectors
with respect to $\Vec P$ and with respect to $\Vec P_h$. However,
this difference can be ignored as far as one neglects subleading
corrections in $P^-$ (i.e., higher twists).

The explicit expression of the symmetric part of the leptonic
tensor in the $\gamma^* N$ frame is \cite{Mulders:1995dh}
\bq
L^{\mu \nu}_{\rm (s)} &=&
\frac{Q^2}{y^2} \left \{
- 2 \left ( 1 - y + \frac{y^2}{2} \right ) g_{\perp}^{\mu \nu}
 + 4 (1 - y) \left [
\frac{x_B^2}{Q^2}\,   p^{\mu} p^{\nu} +
\frac{Q^2}{4 x_B^2}\, n^{\mu} n^{\nu} +
\frac{1}{2}\, p^{\{ \mu} n^{\nu \}} \right ] \right.
\nonumber \\
& & + \left. 4 (1 - y) \left (\hat \ell_{\perp}^{\mu}
\hat \ell_{\perp}^{\nu}  + \frac{1}{2} \, g_{\perp}^{\mu \nu}
\right )
 +  2 (2 - y) \sqrt{1 - y}
\left [ \frac{x_B}{Q} \, p^{\{ \mu} \hat \ell_{\perp}^{\nu \}}
+ \frac{Q}{2 x_B}  \, n^{\{ \mu} \hat \ell_{\perp}^{\nu \}}
\right ] \right \}\,,
\label{symlept}
\eq
where $\ell_{\perp}^{\mu}$ is the transverse component of the incoming
and outgoing lepton momentum ($\hat\ell_{\perp}^{\mu} =
\ell_{\perp}^{\mu} / |\Vec \ell_{\perp}|$),
and $g_{\perp}^{\mu \nu} = g^{\mu \nu} - (p^{\mu} n^{\nu}
+ p^{\nu} n^{\mu})$.
The antisymmetric part of the leptonic tensor
reads ($\lambda_l$ is the helicity of the lepton
and $\epsilon_{\perp}^{\mu \nu}
\equiv \epsilon^{\mu \nu \rho \sigma} p_{\rho} n_{\sigma}$)
\bq
L^{\mu \nu}_{\rm (a)} &=& \frac{Q^2}{y^2} \left \{
- \I \, \lambda_l \, y (2 - y) \, \epsilon_{\perp}^{\mu \nu}
- 2 \I \, \lambda_l \, y \sqrt{1 - y} \, \epsilon^{\mu \nu \rho \sigma}
\left ( \frac{x_B}{Q} \, p_{\rho} - \frac{Q}{2 x_B} \, n_{\rho}
\right ) \hat\ell_{\perp \sigma} \right \} \,.
\label{antlept}
\eq

In the parton model, or equivalently at lowest order in QCD,  the
hadronic tensor in the target fragmentation region is represented by
the handbag diagram of Fig.~\ref{handbag} (right) and reads
(to simplify the presentation, we consider only quarks, the
extension to antiquarks being straightforward):
\bq
W^{\mu \nu} &=& \frac{1}{(2 \pi)^4} \, \sum_a
e_a^2 \, \sum_X \int \frac{\D^3 \Vec P_X}{(2 \pi)^3 \, 2 E_X}
\, \int \frac{\D^4 k}{(2 \pi)^4} \,
\int \frac{\D^4 k'}{(2 \pi)^4} \, 2 \pi \, \delta ({k'}^2)
\times \nonumber \\
& & \, [ \anti{u} (k') \gamma^{\mu} \phi (k, P, P_h)]^*
[\anti{u}(k') \gamma^{\nu} \phi(k, P, P_h)]
\times \nonumber \\
& &
\, ( 2 \pi)^4 \, \delta^4 (P - k - P_h - P_X)
\, (2 \pi)^4 \, \delta^4 (k + q - k') \,,
\label{hadten0}
\eq
where we have introduced the matrix elements of the quark field
between the nucleon and the composite state of the hadron and
the target remnant:
\be
\phi_i (k, P, P_h) \equiv \langle P_h, S_h; X \vert
\psi_i (0) \vert P, S \rangle\,.
\label{matel}
\ee
Let us now define the fracture matrix
$\mathcal{M}$ representing the partonic structure of the nucleon target
when it fragments into the final-state hadron:
\bq
 \mathcal{M}_{ij} (k; P,S; P_h,S_h) &=&
\sum_X \int \frac{\D^3 \Vec P_X}{(2 \pi)^3 2 E_X}
\int \frac{\D^4 \xi}{(2 \pi)^4} \,
\E^{\I k \cdot \xi} \times \nonumber \\
& &
\langle P, S \vert \anti \psi_j (0) \vert P_h,S_h; X \rangle
\langle P_h,S_h; X \vert  \psi_i (\xi) \vert P,S \rangle\,.
\label{fractmat}
\eq
In QCD a Wilson line $\mathcal{W}$ connecting the quark
fields must be inserted  in order to ensure gauge invariance.
The antiquark fracture matrix is obtained from (\ref{fractmat})
by replacing $\anti \psi_j$ with $\psi_i$,  and $\psi_i$
with $\anti \psi_j$. In $\mathcal{M}, \psi$ and $\anti \psi$ we
skip a flavor index $a$.

Using the definition (\ref{fractmat}), the hadronic tensor becomes
\be
W^{\mu \nu} = \sum_a e_a^2 \, \int \frac{\D^4 k}{(2 \pi)^4} \,
2 \pi \, \delta [(k + q )^2] \, {\rm Tr} \, \left [
\mathcal{M} \gamma^{\mu} (\slash k + \slash q) \gamma^{\nu}
\right ]  \,.
\label{hadten}
\ee
The quark momentum can be parametrized as
\be
k^{\mu} =
x \, p^{\mu} + \frac{k^2 + \Vec k_{\perp}^2}{2 x} \,n^{\mu}
+ k_{\perp}^{\mu} \simeq
x \, p^{\mu} + k_{\perp}^{\mu}\,,
\ee
with $x \equiv k^-/P^-$, and the delta function
enforcing the on-shellness
of the struck quark sets $x = x_B$:
\be
\delta [(k + q )^2] = \frac{1}{2 P \cdot q} \,
\delta (x_B - k^-/P^-)\,.
\ee

The most general decomposition of $\mathcal{M}$ in a basis of Dirac
matrices would contain terms proportional to $\one, \gamma^{\mu},
\gamma^{\mu} \gamma_5, \gamma_5, \sigma^{\mu \nu} \gamma_5$.
We are interested in leading-twist fracture functions, i.e. in
terms of $\mathcal{M}$ that are of order $(P^-)^1$.
At this order, only the vector, axial and tensor components
of $\mathcal{M}$ appear~\cite{Barone:2001sp}:
\be
\mathcal{M} = \frac{1}{2} \, ( \mathcal{V}_{\mu} \gamma^{\mu}
+ \mathcal{A}_{\mu}  \gamma_5 \gamma^{\mu}
+ \I \, \mathcal{T}_{\mu \nu}\, \sigma^{\mu \nu} \gamma_5 ) \,,
\label{decmat}
\ee
where the coefficients $\mathcal{V}^{\mu}$, $\mathcal{A}^{\mu}$
and $\mathcal{T}^{\mu \nu}$ contain various combinations of the vectors, 
or pseudo-vectors, $P^{\mu}, P_h^{\mu}, k^{\mu}, S^{\mu}$ and $S_h^{\mu}$.
However, for lepto-production in the TFR, the structure of the quark
current is such that only the vector and axial terms contribute at
any twist. This can be explicitly shown using in (\ref{hadten})
the identity
\be
\gamma^{\mu} \gamma^{\rho} \gamma^{\nu}
= (g^{\mu \rho} g^{\nu \sigma} + g^{\mu \sigma} g^{\nu \rho}
- g^{\mu \nu} g^{\rho \sigma} ) \gamma_{\sigma} +
\I \, \epsilon^{\mu \rho \nu \sigma} \gamma_5 \gamma_{\sigma} \,,
\ee
which allows splitting $W^{\mu \nu}$ into a symmetric and an
antisymmetric part:
\bq
W^{\mu \nu}_{\rm (s)} &=& \frac{1}{2 P \cdot q}
\,  \sum_a e_a^2 \, \int \frac{\D^4 k}{(2 \pi)^3} \,
\delta (x_B - k^-/P^-) \,
 \left [ \Strut
(k^{\mu} + q^{\mu}) \, {\rm Tr} \, (\mathcal{M} \gamma^{\nu})
\right. \nonumber \\
& & \hspace{1cm} \left.
+ \, (k^{\nu} + q^{\nu}) \, {\rm Tr} \, (\mathcal{M} \gamma^{\mu})
- g^{\mu \nu} (k_{\rho} + q_{\rho}) \, {\rm Tr} \, (\mathcal{M}
\gamma^{\rho} )\Strut \right ] 
\label{ws}
\\
W^{\mu \nu}_{\rm (a)} &=& \frac{1}{2 P \cdot q}
\,  \sum_a e_a^2 \, \int \frac{\D^4 k}{(2 \pi)^3} \,
\delta (x_B - k^-/P^-) \, \I
\epsilon^{\mu \rho \nu \sigma}
(k_{\rho} + q_{\rho}) \, {\rm Tr} \, (\mathcal{M} \gamma_{\sigma} \gamma_5)
\,.
\label{wa}
\eq
One then sees that the hadronic tensor
contains
 $\mathcal{V}^{\mu} = \frac{1}{2}
{\rm Tr} \, (\mathcal{M} \gamma^{\mu})$
and $\mathcal{A}^{\mu} =
\frac{1}{2} {\rm Tr} \, (\mathcal{M} \gamma^{\mu} \gamma_5)$
only.
The absence of a tensor term in $\mathcal{M}$
means that single-particle lepto-production does not probe
any fracture function of transversely polarised quarks:
this is easily understood by looking at the handbag diagram for target
fragmentation, which cannot flip the helicity of the struck quark.
In order to observe the transverse polarisation of quarks,
which is described by chirally-odd fracture functions,
one needs to detect a second hadron in the CFR, in coincidence
with the one in the TFR.

\section{Leading-twist polarized
and transverse-momentum dependent fracture functions}

The polarized and transverse-momentum
dependent fracture functions appear in the expansion of the leading twist
projections ($\Gamma = \gamma^-,\gamma^- \gamma_5, \I \sigma^{i-} \gamma_5$)
\bq
& & \mathcal{M}^{[\Gamma]} (x_B, \Vec k_{\perp},
\zeta, \Vec P_{h \perp})
\nonumber \\
& & \hspace{1cm}
\equiv
\frac{1}{4 \zeta} \int \frac{\D k^+ \, \D k^-}{(2 \pi)^3}
\, \delta (k^- - x_B P^-) \, {\rm Tr} \, (\mathcal{M} \, \Gamma)
\nonumber \\
& &
\hspace{1cm}
=
\frac{1}{4 \zeta}
\, \int \frac{\D \xi^+ \,
\D^2 \Vec \xi_{\perp}}{(2 \pi)^6}
\, \E^{\I (x_B P^- \xi^+ - \Vec k_{\perp}
\cdot \Vec \xi_{\perp})} \,
\sum_X \int \frac{\D^3 \Vec P_X}{(2 \pi)^3 \, 2 E_X}
\times \nonumber \\
& & \hspace{1.5cm}
\langle P, S\vert \anti \psi(0) \Gamma
\vert  P_h, S_h; X \rangle \langle  P_h, S_h; X \vert
 \psi(\xi^+, 0, \Vec \xi_{\perp})
\vert P, S \rangle \,.
\label{proj}
\eq
These represent the conditional probabilities to find an unpolarized
($\Gamma = \gamma^-$),  a longitudinally polarized
($\Gamma = \gamma^- \gamma_5$) or a transversely polarized
($\Gamma = \I \sigma^{i-} \gamma_5$) quark with longitudinal momentum
fraction $x_B$ and transverse momentum $\Vec k_{\perp}$ inside a nucleon
fragmenting into a hadron carrying a fraction $\zeta$ of the nucleon
longitudinal momentum and a transverse momentum $\Vec P_{h \perp}$.
Again, in QCD a Wilson line $\mathcal{W}$ must be inserted, which
for $\Vec k_{\perp}$-dependent distributions includes transverse links
and is generally rather complicated \cite{Belitsky:2002sm,Bomhof:2006dp}:
its explicit structure, however, is irrelevant for our purposes.

The most general parameterization of the traced fracture matrix 
(\ref{proj}) can be written as:
\bq
\mathcal{M}^{[\gamma^-]}
&=&
\hat{ M}
+ \frac{\Vec P_{h \perp} \times \Vec S_{\perp}}{m_h} \,
\hat{M}_T^h + \frac{\Vec k_{\perp} \times
\Vec S_{\perp}}{m_N} \, \hat{M}_T^{\perp}
 +
\frac{S_{\parallel} \, (\Vec k_{\perp}
\times \Vec P_{h \perp})}{m_N \, m_h} \,
\hat{M}_L^{\perp h}
\label{v1v2} \\
\mathcal{M}^{[\gamma^- \gamma_5]}
&=&
S_{\parallel} \, \Delta \hat{M}_L
+ \frac{ \Vec P_{h \perp} \cdot \Vec S_{\perp}}{m_h} \,
\Delta \hat{M}_T^h
+ \frac{\Vec k_{\perp} \cdot \Vec S_{\perp}}{m_N}
\, \Delta \hat{M}_T^{\perp}
 +  \frac{\Vec k_{\perp} \times
\Vec P_{h \perp}}{m_N \, m_h} \,
\Delta \hat{M}^{\perp h}
\label{a1a2} \\
\mathcal{M}^{[\I \, \sigma^{i -} \gamma_5]}
&=& S_{\perp}^i \, \Delta_T \hat{M}_T
+ \frac{S_{\parallel} \, P_{h \perp}^i}{m_h} \, \Delta_T \hat{M}_L^h
+ \frac{S_{\parallel} \, k_{\perp}^i}{m_N} \,
\Delta_T \hat{M}_L^{\perp}
\nonumber \\
& & + \, \frac{(\Vec P_{h \perp} \cdot \Vec S_{\perp})
\, P_{h \perp}^i}{m_h^2} \, \Delta_T \hat{M}_T^{hh}
+ \frac{(\Vec k_{\perp} \cdot \Vec S_{\perp})
\, k_{\perp}^i}{m_N^2} \, \Delta_T \hat{M}_T^{\perp \perp}
\nonumber \\
& & + \frac{(\Vec k_{\perp} \cdot \Vec S_{\perp})
\, P_{h \perp}^i - (\Vec P_{h \perp} \cdot \Vec S_{\perp})
\, k_{\perp}^i }{m_N m_h} \, \Delta_T \hat{M}_T^{\perp h}
\nonumber \\
& & + \, \frac{\epsilon_{\perp}^{ij} P_{h \perp j}}{m_h}
\, \Delta_T \hat{M}^h
+ \frac{\epsilon_{\perp}^{ij} k_{\perp j}}{m_N}
\, \Delta_T \hat{M}^{\perp}\,,
\label{t1t2}
\eq
where by the vector product of the two-dimensional vectors we mean the
pseudo-scalar quantity $\Vec a_{\perp} \times \Vec b_{\perp} =
\epsilon_{\perp i j} \, a_{\perp}^i b_{\perp}^ j = \vert
\Vec a _{\perp} \vert \vert \Vec b_{\perp} \vert \, \sin (\phi_b - \phi_a)$.
All fracture functions depend on the scalar variables
$x_B, \Vec k_{\perp}^2, \zeta, \Vec P_{h \perp}^2,
\Vec k_{\perp} \cdot \Vec P_{h \perp}$. An important point to
notice is that while parity invariance constrains the structure of the
fracture matrix, time reversal invariance does not, since
$\mathcal{M}$, similarly to the fragmentation
matrix,  contains the out-states  $\vert P_h, S_h; X \rangle$.
In fact the fracture functions of
Eqs.~(\ref{v1v2})--(\ref{t1t2}) can be seen to reflect the independent
combinations, with the appropriate parity properties, of all vectors
and pseudo-vectors at our disposal.

As most of the functions introduced above appear for the first time,
a few words about them and some explanation of the notations adopted
can be useful. We denote by $\hat{M}$ the unintegrated fracture functions
of unpolarized quarks, by $\Delta \hat{M}$ the unintegrated fracture
functions of longitudinally polarized quarks and by $\Delta_T \hat{M}$
the unintegrated fracture functions of transversely polarized quarks.
The subscripts $L$ and $T$, appended to $\hat{M}$, label the
polarization of the target (no subscript = unpolarized,
$L=$ longitudinally polarized, $T=$ transversely polarized).
The superscripts  $h$ and $\perp$ signal the presence of factors
$P_{h \perp}^i$ and $k_{\perp}^i$, respectively. Fracture functions
integrated over $\Vec k_{\perp}$ will not have the hats.

\begin{itemize}
\item
$\hat{M}$ is the unintegrated distribution of unpolarized quarks
inside an unpolarized  nucleon fragmenting into a spinless hadron
emitted with a non-zero transverse momentum. Integrating
$\hat{M}$ over $\Vec P_{h \perp}$ and $\Vec k_{\perp}$,
one obtains the collinear fracture function
introduced in Refs.~\cite{Trentadue:1993ka,Graudenz:1994dq}:
\be
M(x_B, \zeta) = \int \D^2 \Vec P_{h \perp} \, \int \D^2
\Vec k_{\perp} \, \hat{M} (x_B, \Vec k_{\perp}^2,
\zeta, \Vec P_{h \perp}^2, \Vec k_{\perp} \cdot \Vec P_{h \perp})\,.
\ee

\item
$\hat{M}_T^h$ and $\hat{M}_T^{\perp}$ are new fracture functions
describing the distributions of unpolarized quarks inside a
transversely polarized target. If we integrate
$\mathcal{M}^{[\gamma^-]}$ over the quark transverse momentum
by means of the identities in the Appendix, we are left with two
$\Vec P_{h \perp}$-dependent fracture functions:
\be
\int \D^2 \Vec k_{\perp} \, \mathcal{M}^{[\gamma^-]}
= M (x_B, \zeta, \Vec P_{h \perp}^2)
+ \frac{\Vec P_{h \perp} \times \Vec S_{\perp}}{m_h}
\, M_T^h (x_B, \zeta, \Vec P_{h \perp}^2)\,,
\label{intm}
\ee
where
\be
M(x_B, \zeta, \Vec P_{h \perp}^2)
= \int \D^2 \Vec k_{\perp} \,
\hat{M} (x_B, \Vec k_{\perp}^2, \zeta, \Vec P_{h \perp}^2,
\Vec k_{\perp} \cdot \Vec P_{h \perp})
\label{intm1}
\ee
was called ``extended fracture function'' in Ref.~\cite{Grazzini:1997ih}.
$M_T^h$ is obtained from a combination of two unintegrated fracture
functions:
\be
M_T^h(x_B, \zeta, \Vec P_{h \perp}^2)
= \int \D^2 \Vec k_{\perp} \,
\left \{ \hat{M}_T^h
+ \frac{m_h}{m_N}
\frac{\Vec k_{\perp} \cdot \Vec P_{h \perp}}{\Vec P_{h \perp}^2}
\, \hat{M}_T^{\perp} \right \}\,\cdot
\label{intm2}
\ee
Notice that only the correlation
$\Vec P_{h \perp} \times \Vec S_{\perp} = \vert \Vec P_{h\perp}
\vert \vert \Vec S_{\perp} \vert \, \sin (\phi_S - \phi_h)$
survives upon integration over $\Vec k_{\perp}$.

\item
$\hat{M}_L^{\perp h}$, the last independent fracture function
contributing to $\mathcal{M}^{[\gamma^-]}$, describes the distribution
of unpolarized quarks in a longitudinally polarized nucleon. It can only
exist thanks to the scalar combination $S_{\parallel}\,  (\Vec k_\perp \times
\Vec P_{h \perp})$ and does not survive upon integration.

\item
Turning to $\mathcal{M}^{[\gamma^- \gamma^5]}$, $\Delta \hat{M}_L$ is
the unintegrated fracture function of longitudinally polarized quarks
in a longitudinally polarized target; it yields the helicity fracture
function $\Delta M_L (x_B, \zeta)$, once integrated over
$\Vec k_{\perp}$ and $\Vec P_{h \perp}$.

\item
$\Delta \hat{M}_T^{h}$, $\Delta \hat{M}_T^{\perp}$ and
$\Delta \hat{M}^{\perp h }$ are new fracture functions describing
the distribution of longitudinally polarized quarks inside a
transversely polarized nucleon (the first two) and an unpolarized
nucleon ($\Delta \hat{M}^{\perp h}$). Integrating
$\mathcal{M}^{[\gamma^- \gamma_5]}$ over $\Vec k_{\perp}$ with the
help of the identities in the Appendix yields
\be
\int \D^2 \Vec k_{\perp} \,
\mathcal{M}^{[\gamma^- \gamma_5]}
= S_{\parallel} \, \Delta M_L (x_B, \zeta, \Vec P_{h \perp}^2)
+ \frac{\Vec P_{h \perp} \cdot \Vec S_{\perp}}{m_h}
\, \Delta M_T^h (x_B, \zeta, \Vec P_{h \perp}^2)\,,
\label{intdeltam}
\ee
where $\Delta M_L$ is
\be
\Delta M_L(x_B, \zeta, \Vec P_{h \perp}^2)
= \int \D^2 \Vec k_{\perp} \,
\Delta \hat{M}_L (x_B, \Vec k_{\perp}^2, \zeta, \Vec P_{h \perp}^2,
\Vec k_{\perp} \cdot \Vec P_{h \perp})\,,
\label{intdeltam1}
\ee
and $\Delta M_T^h$ is related to two unintegrated
fracture functions as follows
\be
\Delta M_T^h (x_B, \zeta, \Vec P_{h \perp}^2)
= \int \D^2 \Vec k_{\perp}
\, \left \{ \Delta \hat{M}_T^h +
\frac{m_h}{m_N}
\frac{\Vec k_{\perp} \cdot \Vec P_{h \perp}}{\Vec P_{h \perp}^2}
\, \Delta \hat{M}_T^{\perp} \right \}\,.
\label{intdeltam2}
\ee
In this case, the only surviving angular correlation is
 $\Vec P_{h \perp}
\cdot \Vec S_{\perp} = \vert \Vec P_{h \perp} \vert
\vert \Vec S_{\perp} \vert \, \cos (\phi_S - \phi_h)$.

\item

The fracture functions of transversely polarized quarks,
contained in $\mathcal{M}^{[\I \sigma^{i -} \gamma_5]}$,
are not probed in single-particle SIDIS -- the
process we are interested in here -- and will be discussed
in a separate paper devoted to two-hadron lepto-production\footnote{Some
transversely polarized fracture functions have been considered by
Sivers \cite{Sivers:2008,Sivers:2009qj}, but their transverse momentum
and transverse spin structure has not been explored.}.

\end{itemize}

The inclusive lepto-production of a single spinless hadron in
the TFR involves the integration over $\Vec k_\perp$. It then
probes, Eqs.~(\ref{intm}) and (\ref{intdeltam}), four fracture
functions: $M$, $\Delta M_L$, $M_T^h$, $\Delta M_T^h$.
These four fracture functions, giving the distribution of
final hadrons inside polarized nucleons, can be seen as the analog
of some TMDs, which are defined in the CFR.
$M(x_B, \zeta, \Vec P_{h \perp}^2)$ and
$\Delta M_L(x_B, \zeta, \Vec P_{h \perp}^2)$
correspond, respectively, to the unintegrated unpolarized and
helicity distributions. $M_T^h(x_B, \zeta, \Vec P_{h \perp}^2)$,
which describes the distribution of unpolarized final hadrons $h$
inside a transversely polarized nucleon, is the TFR analog of the
Sivers distribution function $f_{1T}^{\perp}(x_B, \Vec k_{\perp}^2)$
\cite{Sivers:1989cc,Anselmino:1998yz}, which describes the
distribution of unpolarized quarks inside a transversely polarized
nucleon (see Ref.~\cite{Barone:2010zz} for a review of the theoretical
and experimental work on the subject).
$\Delta M_T^h (x_B, \zeta, \Vec P_{h \perp}^2)$ is similar
to the distribution function $g_{1T}^\perp (x_B, \Vec k_{\perp}^2)$,
which describes longitudinally polarized quarks inside a
transversely polarized nucleon
\cite{Ralston:1979,Kotzinian:1995dv,Tangerman:1995hw}.

Two important last general remarks are in order. First of all, the
transverse momenta we are considering, $\Vec k_{\perp}$ and
$\Vec P_{h \perp}$, are intended as {\it intrinsic} momenta, not
generated by gluon radiation (this source of transverse momenta was
explored by Ceccopieri and
Trentadue~\cite{Ceccopieri:2005zz,Ceccopieri:2007ek}). In other terms,
the effects we are studying are of non-perturbative origin.
Second, while the factorization theorem for collinear and
$\Vec P_{h \perp}$-dependent fracture functions has been
proven~\cite{Grazzini:1997ih,Collins:1997sr}, no analogous result exists
for fully unintegrated fracture functions. In this case, factorization
is just an assumption.

\section{Momentum sum rules}

It is known \cite{Trentadue:1993ka}
that the unpolarized collinear
fracture function $M(x_B, \zeta)$
satisfies a momentum sum rule:
\be
\sum_h \, \int_0^{1 - x_B} \D \zeta \, \zeta \, M(x_B, \zeta) =
(1 - x_B) \, f_1(x_B)\,,
\label{sr0}
\ee
where $\sum_h$ is a sum over all hadrons, and $f_1$
is the ordinary number density of quarks. Equation~(\ref{sr0})
is easily understood by
remembering the probabilistic
interpretation of $M$ and
noticing that, if the target
emits a quark with momentum fraction $x_B$, the
total momentum fraction available for the
final hadron is $1 - x_B$.

A set of generalized sum rules for the transverse-momentum dependent
fracture functions introduced in the previous Section can be derived
as follows. First of all, the integral over $\Vec P_X$ and the $\sum_X$
of the out-states $\vert P_h, S_h; X \rangle$ yields the number
operator of the hadrons $h$ \cite{Collins:1982,Levelt:1992a1}:
\be
\sum_X \int \frac{\D^3 \Vec P_X}{(2 \pi)^3 \, 2 E_X}
\, \vert P_h, S_h; X \rangle \langle
P_h, S_h; X \vert = a_h^{\dagger} a_h \,,
\label{nh}
\ee
where $a_h$ ($a_h^{\dagger}$) is the annihilation
(creation) operator of the hadron $h$ with momentum $P_h$.
The traced fracture matrix (\ref{proj}) can thus be
rewritten as
\be
\mathcal{M}^{[\Gamma ]} =
\frac{1}{4 \zeta}
\, \int \frac{\D \xi^+ \,
\D^2 \Vec \xi_{\perp}}{(2 \pi)^6}
\, \E^{\I (x_B P^- \xi^+ - \Vec k_{\perp}
\cdot \Vec \xi_{\perp})} \,
\langle P, S\vert \anti \psi(0) \, \Gamma \, a_h^{\dagger} \, a_h
\, \psi(\xi^+, 0, \Vec \xi_{\perp})
\vert P, S \rangle \,.
\label{proj_bis}
\ee
If we now multiply $\mathcal{M}^{[\Gamma]}$
by $\zeta$, integrate over $\zeta$ and $\Vec P_{h \perp}$,
 and sum over all
hadrons, we get
\bq
& & \sum_h \, \int_0^{1 - x_B} \D \zeta
\, \zeta \, \int \D^2 \Vec P_{h \perp}
\, \mathcal{M}^{[\Gamma]} (x_B, \Vec k_{\perp},
\zeta, \Vec P_{h \perp})
\nonumber \\
& & \hspace{1cm}
= \sum_h \int_0^{1 - x_B} \D \left ( \frac{P_h^-}{P^-}
\right ) \, \frac{P_h^-}{P^-}
\, \int \D^2 \Vec P_{h \perp} \,
\mathcal{M}^{[\Gamma]} (x_B, \Vec k_{\perp},
\zeta, \Vec P_{h \perp})
\nonumber \\
&& \hspace{1cm} =
\frac{1}{2} \int \frac{\D \xi^+ \D^2 \Vec \xi_{\perp}}{(2 \pi)^3}
\, \E^{\I (x_B P^- \xi^+ - \Vec k_{\perp} \cdot \Vec \xi_{\perp})}
\sum_h \int \frac{\D P_h^- \, \D^2 \Vec P_{h \perp}}{(2 \pi)^3 \, 2 P_h^-}
\, \langle P, S \vert \anti{\psi} (0)  \, \Gamma
\, \frac{P_h^-}{P^-} \, a_h^{\dagger} a_h \, \psi (\xi^+, 0,
\Vec \xi_{\perp}) \vert P, S \rangle
\nonumber \\
&& \hspace{1cm} = (1 - x_B) \,
\Phi^{[\Gamma]} (x_B, \Vec k_{\perp})\, ,
\label{sr1}
\eq
where
\be
\Phi^{[\Gamma]}(x_B, \Vec k_{\perp}) =
\frac{1}{2} \int \frac{\D \xi^+ \, \D^2 \Vec \xi_{\perp}}{(2 \pi)^3}
\, \E^{\I (x_B P^- \xi^+ - \Vec k_{\perp} \cdot \Vec \xi_{\perp})}
\, \langle P, S \vert \anti \psi (0) \,
\Gamma \, \psi (\xi^+, 0, \Vec \xi_{\perp}) \vert P, S \rangle
\ee
is the traced quark correlation matrix containing the
transverse momentum dependendent distributions.
Equation~(\ref{sr1}) establishes a relation between the
fracture functions of Eqs.~(\ref{v1v2})-(\ref{t1t2})
and the TMDs appearing in the expansion of
$\Phi^{[\gamma^-]}$, $\Phi^{[\gamma^- \gamma_5]}$
and $\Phi^{[\I \sigma^{i -} \gamma_5]}$
(for which see, e.g., Ref.~\cite{Barone:2010zz}).
The final results, obtained by using the identities of the Appendix,
are
\bq
& & \sum_h \int \D \zeta \, \zeta \, \int \D^2 \Vec P_{h \perp} \,
\hat{M} = (1 - x_B) \, f_1 (x_B, \Vec k_{\perp}^2) 
\label{f1} \\
& & \sum_h \int \D \zeta \, \zeta \, \int \D^2 \Vec P_{h \perp}
\left \{
\hat{M}_T^{\perp} + \frac{m_N}{m_h}
\, \frac{\Vec k_{\perp} \cdot \Vec P_{h \perp}}{\Vec
k_{\perp}^2} \, \hat{M}_T^h \right \}
= - (1 - x_B) \, f_{1T}^{\perp} (x_B, \Vec k_{\perp}^2)
\label{f1T}  \\
& & \sum_h \int \D \zeta \, \zeta \, \int \D^2 \Vec P_{h \perp}
\Delta \hat{M}_L = (1 - x_B) \, g_{1L} (x_B, \Vec k_{\perp}^2)
\label{g1L} \\
& & \sum_h \int \D \zeta \, \zeta \, \int \D^2 \Vec P_{h \perp}
\left \{ \Delta \hat{M}_T^{\perp} +
\frac{m_N}{m_h} \, \frac{\Vec k_{\perp} \cdot \Vec P_{h \perp}}{\Vec
k_{\perp}^2} \, \Delta \hat{M}_T^h \right \} =
(1 - x_B) \, g_{1T}(x_B, \Vec k_{\perp}^2) 
\label{g1T} \\
& & \sum_h \int \D \zeta \, \zeta \, \int \D^2 \Vec P_{h \perp}
\left \{ \Delta_T \hat{M}_L^{\perp} +
\frac{m_N}{m_h} \, \frac{\Vec k_{\perp} \cdot \Vec P_{h \perp}}{\Vec
k_{\perp}^2} \, \Delta_T \hat{M}_L^h \right \} =
(1 - x_B) \, h_{1L}^{\perp}(x_B, \Vec k_{\perp}^2) 
\label{h1L} \\
& &
\sum_h \int \D \zeta \, \zeta \, \int \D^2 \Vec P_{h \perp}
\left \{ \Delta_T \hat{M}^{\perp} +
\frac{m_N}{m_h} \, \frac{\Vec k_{\perp} \cdot \Vec P_{h \perp}}{\Vec
k_{\perp}^2} \, \Delta_T \hat{M}^h \right \}
= - (1 - x_B) \, h_{1}^{\perp}(x_B, \Vec k_{\perp}^2) 
\label{h1perp} \\
& &
\sum_h \int \D \zeta \, \zeta \, \int \D^2 \Vec P_{h \perp}
\left \{ \Delta_T \hat{M}_T^{\perp \perp} 
+ \frac{m_N^2}{m_h^2}\frac{
2 (\Vec k_{\perp} \cdot \Vec P_{h \perp})^2
- \Vec k_{\perp}^2 \Vec P_{h \perp}^2}{(\Vec k_{\perp}^2)^2}
 \, \Delta_T \hat{M}_T^{hh}
\right \}  \\
\nonumber
& & \hspace{1cm} =
(1 - x_B) \, h_{1T}^{\perp}(x_B, \Vec k_{\perp}^2) 
\label{h1Tperp} \\
& &
\sum_h \int \D \zeta \, \zeta \, \int \D^2 \Vec P_{h \perp}
\left \{ \Delta_T \hat{M}_T + \frac{\Vec k_{\perp}^2}{2m_N^2} \,
\Delta_T \hat{M}_T^{\perp\perp} + \frac{\Vec P_{h\perp}^2}{2m_h^2} \,
\Delta_T \hat{M}_T^{hh} \right \}
= (1 - x_B) \, h_{1}(x_B, \Vec k_{\perp}^2) \,.
\label{h1} 
\eq
These are the momentum sum rules satisfied by the unintegrated
fracture functions. They might be useful for constraining and guiding
simple models of fracture functions.

\section{Cross sections and angular distributions}

Contracting the hadronic tensor, Eqs.~(\ref{ws}, \ref{wa}),
with the symmetric and antisymmetric part of the leptonic tensor, 
Eqs.~(\ref{symlept}, \ref{antlept}), and using Eq.~(\ref{proj}), yields
\bq
L_{\rm (s)}^{\mu \nu} W_{\mu \nu}^{\rm (s)}
&=& \frac{8 Q^2}{y^2} \left (1 - y + \frac{y^2}{2}
\right ) \, \zeta \, \sum_a e_a^2 \,
\int \D^2 \Vec k_{\perp} \, \mathcal{M}^{[\gamma^-]}
\label{contr_s} \\
L_{\rm (a)}^{\mu \nu} W_{\mu \nu}^{\rm (a)}
&=& \lambda_l \frac{8 Q^2}{y^2} \, y \left (1 - \frac{y}{2} 
\right ) \, \zeta \, \sum_a e_a^2 \,
\int \D^2 \Vec k_{\perp} \, \mathcal{M}^{[\gamma^- \gamma_5]}\,.
\label{contr_a}
\eq

We focus on three processes:

\begin{enumerate}

\item
{\it lepto-production of a spinless hadron}, $l + N \to l' + h + X$;

\item
{\it lepto-production of a spinless hadron plus a quark jet},
$l + N \to l' + h + {\rm jet} + X$;

\item
{\it lepto-production of a polarized hadron}, $l + N \to
l' + h^{\uparrow} + X$ (integrated over all
transverse momenta).

\end{enumerate}

\subsection{Lepto-production of a spinless hadron}

Consider the lepto-production of an unpolarized or spinless hadron
(for instance, pion lepto-production, which is the most common process).
Inserting Eqs.~(\ref{intm}, \ref{intdeltam}) into 
Eqs.~(\ref{contr_s}, \ref{contr_a}), and using Eq.~(\ref{sidis9}),
one finds that the cross section for this process is
\bq
& & \frac{\D \sigma^{\rm TFR}}{\D x_B \, \D y \, \D \zeta\,
\D^2 \Vec P_{h \perp} \, \D \phi_S}
 =
\frac{2 \alpha_{\rm em}^2}{Q^2 y} \,
\left \{
\left (1 - y + \frac{y^2}{2} \right )
\right.
\nonumber \\
& & \hspace{1cm}
\times \, \sum_a e_a^2 \,
 \left  [   M(x_B, \zeta, \Vec P_{h \perp}^2)
-   \vert \Vec S_{\perp} \vert \, \frac{\vert \Vec P_{h \perp}\vert}{m_h}
\, M_{T}^h (x_B, \zeta, \Vec P_{h \perp}^2) \, \sin (\phi_h - \phi_S)
\right ]
\nonumber \\
& &  \hspace{1cm} + \,
\lambda_l \, y \, \left (1 - \frac{y}{2}\right )
\sum_a e_a^2 \,
 \left [ \STRUT
S_{\parallel} \, \Delta M_{L} (x_B, \zeta, \Vec P_{h \perp}^2)
\right.
\nonumber \\
& & \hspace{1cm}
+ \, \left. \left.
\vert \Vec S_{\perp} \vert \, \frac{\vert \Vec P_{h \perp} \vert}{m_h}
\, \Delta M_{T}^h (x_B, \zeta, \Vec P_{h \perp}^2) \, \cos (\phi_h - \phi_S)
\right ] \right \}\,.
\label{cross1}
\eq
As already mentioned, this process is described by four fracture functions,
two of which appear when the target is transversely polarized.
The only azimuthal modulations arise from correlations between the
transverse spin of the target and the transverse momentum of the final
hadron, and are of the type $\sin (\phi_h - \phi_S)$ and 
$\cos (\phi_h - \phi_S)$: the former manifests itself when the lepton beam
is unpolarized, the latter, when the beam is longitudinally polarized.

It is interesting to compare our result with the corresponding situation
in the current fragmentation region, where at leading twist there are
six different modulations: besides the two which also appear in the TFR,
that is $\sin (\phi_h - \phi_S)$ and $\cos (\phi_h - \phi_S)$,
there is a $\cos 2 \phi_h$ term associated with the
Boer-Mulders effect in unpolarized SIDIS
\cite{Boer:1997nt,Barone:2008tn,Barone:2009hw},
a $\sin 2 \phi_h$ term for a longitudinally polarized target,
and, in the transversely polarized case, the $\sin (\phi_h + \phi_S)$
term, which gives access to the transversity
distribution~\cite{Collins:1992kk,Anselmino:2007fs},
and the $\sin (3 \phi_h - \phi_S)$ term involving the
distribution $h_{1T}^{\perp}$. All these additional asymmetries
arise due to the Collins effect in the fragmentation of a transversely
polarized quark. The absence of these type of angular distributions in
the TFR might have some relevance in phenomenological analyses.

Referring to the general parametrization of the SIDIS cross section
presented in Ref.~\cite{Bacchetta:2006tn},
$M_T^h$ and $\Delta M_T^h$ in (\ref{cross1}) contribute, in the TFR,
to the structure functions $F_{UT,T}^{\sin (\phi_h - \phi_S)}$
and $F_{LT}^{\cos (\phi_h - \phi_S)}$, respectively:
\bq
& &  \left[ F_{UT,T}^{\sin (\phi_h - \phi_S)} \right]_{\rm TFR}
=  - \sum_a e_a^2
\, x_B  \, \frac{\vert \Vec P_{h \perp} \vert}{m_h}
\, M_{T}^h (x_B, \zeta, \Vec P_{h \perp}^2)
\label{fut} \\
 & & \left[ F_{LT}^{\cos (\phi_h - \phi_S)} \right]_{\rm TFR}
=  \sum_a e_a^2
\, x_B  \, \frac{\vert \Vec P_{h \perp} \vert}{m_h}
\, \Delta M_{T}^h (x_B, \zeta, \Vec P_{h \perp}^2)\,.
\label{flt}
\eq

For comparison, we recall that in the CFR the two structure functions
above are given by \cite{Bacchetta:2006tn}:
\be
\left[ F_{UT,T}^{\sin (\phi_h - \phi_S)} \right]_{\rm CFR}
= \mathcal{C} \, \left [ - \frac{\hat {\Vec h} \cdot
\Vec k_{\perp}}{m_N} \, f_{1T}^{\perp} \, D_1
\right ] \,, \;\;\;\;\;\;
\left[ F_{LT}^{\cos (\phi_h - \phi_S)} \right]_{\rm CFR}
= \mathcal{C} \, \left [  \frac{\hat {\Vec h} \cdot
\Vec k_{\perp}}{m_N} \, g_{1T} \, D_1
\right ] \,,
\label{flt_cfr}
\ee
where $\hat{\Vec h} \equiv \Vec P_{h \perp}/\vert \Vec P_{h \perp}\vert$
and $\mathcal{C}$ denotes the transverse momentum convolution
\bq
\mathcal{C} \, [w \, f \, D]
&\equiv& \sum_a e_a^2 \, x_B \int \D^2 \Vec k_{\perp}
\int \D^2 \Vec \kappa_{\perp}
\, \delta^2 (\Vec k_{\perp} - \Vec \kappa_{\perp}
- \Vec P_{h \perp}/z ) \nonumber \\
& & \hspace{1cm} \times \,  w(\Vec k_{\perp}, \Vec \kappa_{\perp})
\, f(x_B, \Vec k_{\perp}^2) \, D(z_h, \Vec \kappa_{\perp}^2)\,.
\label{conv}
\eq

\subsection{Lepto-production of a spinless hadron and a current jet}

Suppose now that the current quark jet is observed in coincidence
with a spinless hadron in the TFR. In this case the
cross section is differential both in $\Vec P_{h \perp}$
and in $\Vec k_{\perp}$. Thus, there is no $\Vec k_{\perp}$
integration in Eqs.~(\ref{contr_s}, \ref{contr_a}).
The final expression for the cross section is
(for simplicity we omit the dependence of fracture functions
on the variables $x_B, \Vec k_{\perp}^2, \zeta, \Vec P_{h \perp}^2,
\Vec k_{\perp} \cdot \Vec P_{h \perp}$)
\bq
& & \frac{\D \sigma^{\rm TFR}}{\D x_B \, \D y \, \D \zeta\,
 \D^2 \Vec P_{h \perp}
\, \D^2 \Vec k_{\perp} \, \D \phi_S}
 =
\frac{2 \alpha_{\rm em}^2}{Q^2 y} \,
\left \{
\left (1 - y + \frac{y^2}{2} \right )
\right.
\nonumber \\
& & \hspace{0.5cm}
\times \, \sum_a e_a^2
\left  [  \hat{M} + S_{\parallel} \, \frac{\vert \Vec P_{h \perp} \vert
\vert \Vec k_{\perp} \vert }{m_h \, m_N} \, \hat{M}_{L}^{\perp h}
\, \sin (\phi_h - \phi_j)
 \right. \nonumber \\
& & \hspace{0.5cm} \left.
-  \vert \Vec S_{\perp} \vert \, \frac{\vert \Vec P_{h \perp}\vert}{m_h}
\, \hat{M}_{T}^h  \, \sin (\phi_h - \phi_S)
-  \vert \Vec S_{\perp} \vert \, \frac{\vert \Vec k_{\perp}\vert}{m_N}
\, \hat{M}_{T}^{\perp}  \, \sin (\phi_j - \phi_S)
\right ]
\nonumber \\
& &  \hspace{0.5cm} +
\lambda_l \, y \, \left (1 - \frac{y}{2}\right )
\sum_a e_a^2 \,
 \left [
\frac{\vert \Vec P_{h \perp} \vert \vert \Vec k_{\perp} \vert}{m_h
\, m_N} \, \Delta \hat{M}^{\perp h}  \, \sin (\phi_h - \phi_j)
+   S_{\parallel} \, \Delta \hat{M}_{L} \right.
\nonumber \\
& & \hspace{0.5cm} +
\left. \left.
\vert \Vec S_{\perp} \vert \, \frac{\vert \Vec P_{h \perp} \vert}{m_h}
\, \Delta \hat{M}_{T}^h \, \cos (\phi_h - \phi_S)
+ \vert \Vec S_{\perp} \vert \, \frac{\vert \Vec k_{\perp} \vert}{m_N}
\, \Delta \hat{M}_{T}^{\perp} \, \cos (\phi_j - \phi_S)
\right ] \right \},
\label{cross2}
\eq
where $\phi_j$ is the azimuthal angle of the jet. The situation is 
richer than in the previous process, but even in this case some of the
modulations appearing in the production of a single hadron in the CFR,
e.g. $\sin (\phi_h + \phi_S)$, are absent since we do not ``measure" the
final quark transverse polarization. From an experimental viewpoint,
the main practical problem is that it is rather difficult to detect
the quark jet.

\subsection{Lepto-production of a polarized hadron integrated over
transverse momenta}

We have assumed so far that the final hadron is spinless or unpolarized.
Relaxing this condition, the panorama of unintegrated fracture functions
becomes extremely complicated. However, the collinear case, that is,
the lepto-production of a polarized hadron integrated over all
transverse momenta, is still manageable. Considering only the leading 
terms in the expansion of the traced fracture matrix we have (the
superscript denotes the polarization state of the final hadron)
\bq
\int \D^2 \Vec P_{h \perp} \, \int \D^2 \Vec k_{\perp}
\, \mathcal{M}^{[\gamma^-]}
\!\!\!&=&\!\!\!
 M(x_B, \zeta)
+ S_{\parallel} \, S_{h \parallel} \, M_L^L (x_B, \zeta)
+ (\Vec S_{\perp} \cdot \Vec S_{h \perp})\,
M_T^{T} (x_B, \zeta) 
\label{v1v8} \\
\int \D^2 \Vec P_{h \perp}  \int \D^2 \Vec k_{\perp} \,
\mathcal{M}^{[\gamma^- \gamma_5]}
\!\!\!&=&\!\!\!
S_{\parallel} \, \Delta M_L(x_B, \zeta)
+ S_{h \parallel} \, \Delta M^L (x_B, \zeta)
+
(\Vec S_{\perp} \times \Vec S_{h \perp}) \,
 \Delta M_T^T  (x_B, \zeta) \,.
\label{a1a8}
\eq
We see that an unpolarized target can emit a longitudinally
polarized quark and a longitudinally polarized hadron, via
$\Delta M^L$, a longitudinally polarized target can emit
an unpolarized quark and a longitudinally polarized
hadron, via $M_L^L$,
and a transversely polarized target
can produce a transversely polarized hadron in two different
ways: via a correlation of the type $\Vec S_{\perp} \cdot \Vec S_{h \perp}$
by emitting an unpolarized quark, or via a correlation
of the type $\Vec S_{\perp} \times \Vec S_{h \perp}$
by emitting a longitudinally polarized quark.

From Eqs. (\ref{v1v8}, \ref{a1a8}) and (\ref{contr_s}, \ref{contr_a}) we
derive the cross section for the process $l \, N \, \to
l' \, h^{\uparrow} \, X$:
\bq
& & \frac{\D \sigma^{\rm TFR}}{\D x_B \, \D y \, \D \zeta\,
\, \D \phi_S \,\D \phi_{S_h}}
 = \frac{\alpha_{\rm em}^2}{\pi \, Q^2 \, y} \,
\left \{ \left (1 - y + \frac{y^2}{2} \right ) \right.
\nonumber \\
& & \hspace{1cm} \times \,
\sum_a e_a^2 \,
 \left  [ \Strut  M(x_B, \zeta)
+  S_{\parallel} \, S_{h \parallel} \, M_{L}^L (x_B, \zeta)
+  \vert \Vec S_{\perp} \vert \,
\vert \Vec S_{h \perp} \vert
\, M_{T}^T (x_B, \zeta)  \, \cos (\phi_{S_h} - \phi_{S})
\right ]
\nonumber \\
& &  \hspace{1cm} + \,
\lambda_l \, y \, \left (1 - \frac{y}{2}\right )
\sum_a e_a^2 \,
 \left [ \Strut
S_{\parallel} \, \Delta M_{L} (x_B, \zeta) +
S_{h \parallel} \, \Delta M^L (x_B, \zeta) \right.
\nonumber \\
& & \hspace{1cm}
+ \left. \left.
\vert \Vec S_{\perp} \vert \, \vert \Vec S_{h \perp} \vert
\, \Delta M_{T}^T (x_B, \zeta)  \, \sin (\phi_{S_h} - \phi_{S}) \Strut
\right ] \STRUT \right \}\,.
\label{cross3}
\eq
The transverse polarization terms
contain modulations of the type $\cos (\phi_{S_h} - \phi_{S})$
and $\sin (\phi_{S_h} - \phi_{S})$, the former involving unpolarized
quarks, the latter longitudinally polarized ones.
It is instructive to compare this result with the
corresponding cross section for current fragmentation, which is
\bq
& & \frac{\D \sigma^{\rm CFR}}{\D x_B \, \D y \, \D z_h\,
\, \D \phi_S \, \D \phi_{S_h}}
 = \frac{\alpha_{\rm em}^2}{\pi \, Q^2 \, y}
\nonumber \\
& & \hspace{0.3cm}
\cdot
\left \{
\left (1 - y + \frac{y^2}{2} \right )
\sum_a e_a^2 \,
 \left  [ \Strut f_{1}(x_B) \, D_{1}(z_h)
+  S_{\parallel} \, S_{h \parallel} \, g_{1}(x_B) G_{1}(z_h) \right ]
\right. \nonumber \\
& & \hspace{0.8cm} - \STRUT (1 - y) \, \vert \Vec S_{\perp} \vert \,
\vert \Vec S_{h \perp} \vert
\, \sum_a e_a^2 \, h_{1}(x_B) H_{1} (z_h)  \, \cos (\phi_S + \phi_{S_h})
\nonumber \\
& &  \hspace{0.8cm} + \left.
\lambda_l \, y \, \left (1 - \frac{y}{2}\right )
\sum_a e_a^2 \,
 \left [ \Strut
S_{\parallel} \, g_1(x_B) D_{1}(z_h) +
S_{h \parallel} \, f_{1}(x_B) G_{1}(z_h) \right ] \right \},
\label{cross3_cfr}
\eq
where $f_1$, $g_1$ and $h_1$ are the unpolarized, helicity and transversity
distributions respectively, and $D_1$, $G_1$, $H_1$ are the corresponding
fragmentation functions. In the CFR cross
section, the transverse spin term comes from a tensor component
of the quark correlation matrix and involves the transversity distribution.

\section{Conclusions and perspectives}

We have presented the formalism of polarized and transverse-momentum
dependent fracture functions to describe SIDIS processes in the TFR.
At leading twist, neglecting all contributions of order 1/$Q$, and
considering only spinless or unpolarized final hadrons, we find 
sixteen fracture functions defined in Eqs.~(\ref{v1v2})--(\ref{t1t2}).
They represent the combined distributions of initial quarks and final
hadrons inside a nucleon; the quark takes part in the elementary
interaction, carrying away (in the collinear limit) a fraction $x_B$
of the nucleon energy $E$, while the remnant of the nucleon, with
energy $(1 - x_B)E$ fragments into the observed final hadron $h$ with
energy $\zeta E = z(1-x_B)E$.
We have taken into account the transverse momenta
of quarks, $\Vec k_\perp$, and of the final hadron $\Vec P_{h\perp}$,
as well as the nucleon and quark polarization. The final hadron
polarization has only been briefly discussed in the collinear, or
fully integrated, case. The fracture functions are divided into
three classes, referring, respectively, to unpolarized,
longitudinally polarized and transversely polarized quarks.

These quantities are probed in the target fragmentation region of
SIDIS. We have seen that when a single hadron is detected in the TFR
only two classes of fracture functions can be measured,
representing  the distributions of unpolarized and longitudinally
polarized quarks. The fracture functions related to the distribution
of transversely polarized quarks are chiral-odd quantities and could
only be accessed by coupling such fracture funtions to other chiral-odd
functions, like the Collins fragmentation function. This could be done
in SIDIS processes with the detection of two final hadrons, one in the
TFR and one in the CFR and will be discussed in a separate paper.

The explicit parton-model expressions of the cross sections
 and angular distributions have been worked out in three cases.
 The first is the
lepto-production of a single spinless hadron in the TFR: in such a case
no information can remain on $\Vec k_\perp$, which is integrated, and
one remains with four $\Vec P_{h \perp}$-dependent fracture functions.
Two of them survive upon integration over $\Vec P_{h \perp}$,
leading to the ordinary, fully integrated, fracture functions (the
unpolarized and the helicity fracture functions). The other two
generate interesting $\sin (\phi_h - \phi_S)$ and
$\cos (\phi_h - \phi_S)$ azimuthal modulations. These fracture functions
are analogous to the Sivers and the $g_{1T}^\perp$ TMDs, which, in the
CFR, indeed generate the same azimuthal dependences. A careful separation
of the TFR and the CFR, when analyzing such azimuthal asymmetries
is crucial.

The second case considered requires that the current jet is detected,
in addition to the hadron in the TFR. In such a case $\Vec k_\perp$
is observed and the eight fracture functions of the first two classes
are fully involved, having all possibilities of correlations between
the transverse momenta (of the jet and of the hadron) and the target
spin.

Finally, we have considered the lepto-production of a polarized hadron
integrated (for obvious reasons of simplicity) over all transverse
momenta. Such a process probes six fracture functions, four of which
require the measurement of the final hadron polarization (and are
different from those previously discussed).

Many possible further developments of this work can be envisaged.
As already stated, the most obvious extension is the full study
of the case in which two hadrons are produced, one in the TFR and
one in the CFR. The handbag parton model diagram for such a
process is shown in Fig. 2 (left). In the simple case in which both
final hadrons are spinless it couples the full set of 
fracture functions to the two leading-twist fragmentation functions
(the unpolarized and the Collins fragmentation functions). The
chiral-odd nature of the Collins function allows to access the
(chiral-odd) fracture functions describing the distributions of
transversely polarized quarks, Eq.~(\ref{t1t2}).

Another potentially very interesting process is, in $N-N$ interactions,
the usual Drell-Yan production of a lepton pair, with, in addition,
the detection of a hadron produced in one of two TFRs, as shown in
Fig. 2 (right). Such a process couples the fracture functions
with the collinear PDFs or the TMDs; for $p-p$ interactions one could
consider the fracture functions for quarks and the CFR distribution
functions for anti-quarks (in a proton). Even better, with $N-\bar p$
interactions one could only refer to quark distributions inside a
proton.

Finally, but not less important, there is the whole field of the
phenomenological applications of the formalism of polarized fracture
functions. This requires modeling in a simple way the fracture functions
and computing the relevant observables in the kinematics of the available
SIDIS experiments (HERMES, COMPASS, JLAB). In particular, it is important
to assess the role of the fracture functions in the measured Sivers
asymmetries, which so much have influenced the exploration of the
3-dimensional momentum structure of the nucleon.

Some phenomenological studies to describe spin phenomena for single hadron
production in the TFR of SIDIS were performed using Monte Carlo event
generators. For example, in Refs.~\cite{Ellis:1995fc,Ellis:2002zv} the
longitudinal polarization of $\Lambda$~baryons produced in the TFR of
SIDIS of polarized leptons off an unpolarized or longitudinally polarized
target was modeled for the transverse momentum integrated case (as in
Sec. 5.3). Some predictions for the Sivers-type modulation of unpolarized 
hadron production in the TFR of SIDIS off transversely polarized protons 
were presented in Ref.~\cite{Kotzinian:2005zg}. However, these results
were obtained by an event generator based on the Lund string model, 
which covers both the $x_F < 0$ and the $x_F > 0$ regions, but does not
distinguish between the two different dynamical mechanisms of fracture
functions and current fragmentation functions.   

The fracture function formalism we considered here gives, for single hadron
production in the TFR, clear and rather simple predictions for azimuthal
asymmetries, Eq.~(\ref{cross1}). The eventual experimental observation
of other azimuthal modulations, for example a Collins-type 
$\sin(\phi_h + \phi_S)$ dependence, would indicate that the QCD fracture 
function approach, with a dynamical separation between the TFR and the 
CFR, does not hold and that long range correlations between the struck 
quark and the produced hadron might be important in the TFR as well.

The ideal experiments to test the fracture function factorization
and measure these new functions in SIDIS, are those being discussed in
the international community and planned at future Electron Ion or
Electron Nucleon Colliders (EIC/ENC). The collider mode and the high
energy would allow a clean kinematical separation between the TFR and
the CFR and a thorough systematic and independent investigation of
the two regions.

\begin{figure}[t]
\begin{center}
\includegraphics[width=0.40\textwidth]
{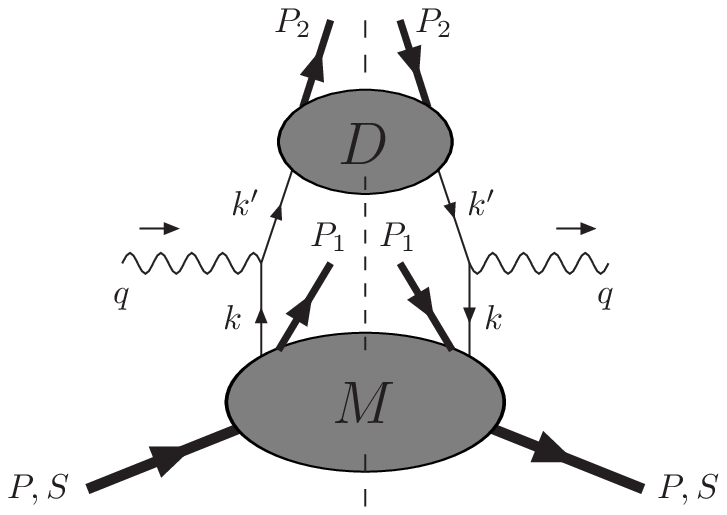}
\hspace{0.2cm}
\includegraphics[width=0.45\textwidth]
{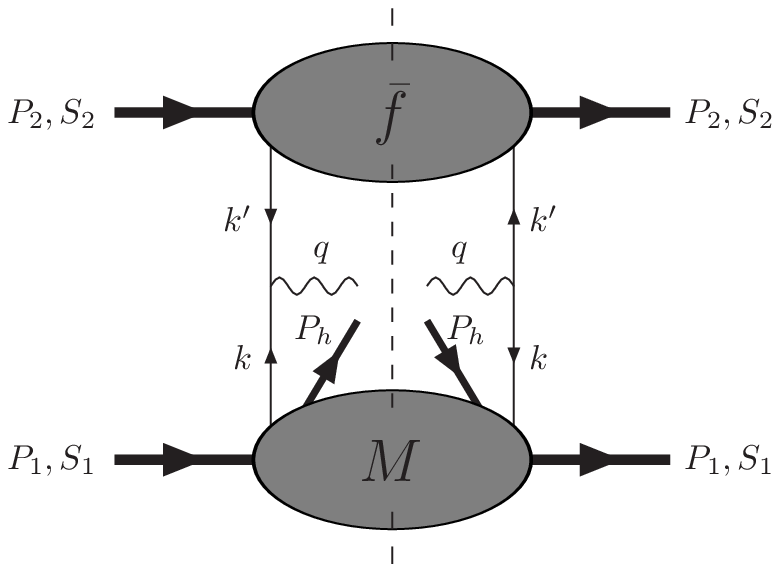}
\caption{The handbag diagram for the lepto-production
of two hadrons, one in the CFR and one in the TFR (left)
and for the Drell-Yan production of a leading hadron (right).}
\label{handbag_two}
\end{center}
\end{figure}

\section*{Acknowledgements}
We thank Markus Diehl for useful discussions and for pointing out a 
redundancy in the list of 
fracture functions appearing in an earlier version of this manuscript. 
We acknowledge partial support by the Italian Ministry of Education,
University and Research (MIUR) in the framework of a Research Project
of National Interest (PRIN 2008).
We also acknowledge partial support by the European Community - Research
Infrastructure Activity under the FP7 program (HadronPhysics2, Grant
agreement 227431), by the Helmholtz Association through
funds provided to the Virtual Institute ``Spin and Strong QCD''(VH-VI-231)
and by Regione Piemonte.

\section*{Appendix}

We exploit the following identities to perform the integration
over $\Vec k_{\perp}$:
\bq
& & \int \D^2 \Vec k_{\perp} \, k_{\perp}^i \, f( \Vec k_{\perp}^2,
\Vec P_{h \perp}^2, \Vec k_{\perp} \cdot \Vec P_{h \perp})
= P_{h \perp}^i \, I_1^{[f]}(\Vec P_{h \perp}^2) \,,
\label{int1} \\
& & \int \D^2 \Vec k_{\perp} \, k_{\perp}^i \, k_{\perp}^j
\, f( \Vec k_{\perp}^2,
\Vec P_{h \perp}^2, \Vec k_{\perp} \cdot \Vec P_{h \perp})
= P_{h \perp}^i \,P_{h \perp}^j \,
 I_2^{[f]}(\Vec P_{h \perp}^2) + g^{ij} \, I_3^{[f]}(\Vec P_{h \perp}^2)\,,
\label{int2}
\eq
with
\bq
I_1^{[f]}(\Vec P_{h \perp}^2)
&=& \int \D^2 \Vec k_{\perp} \,
\frac{\Vec k_{\perp} \cdot \Vec P_{h \perp}}{\Vec P_{h \perp}^2}
\, f(\Vec k_{\perp}^2, \Vec P_{h \perp}^2, \Vec k_{\perp}
\cdot \Vec P_{h \perp})\,,
\label{i1} \\
I_2^{[f]}(\Vec P_{h \perp}^2)
&=& \int \D^2 \Vec k_{\perp} \,
\frac{2 (\Vec k_{\perp} \cdot \Vec P_{h \perp})^2 -
\Vec P_{h \perp}^2 \Vec k_{\perp}^2}{(\Vec P_{h \perp}^2)^2}
\, f(\Vec k_{\perp}^2, \Vec P_{h \perp}^2, \Vec k_{\perp}
\cdot \Vec P_{h \perp})\,,
\label{i2} \\
I_3^{[f]}(\Vec P_{h \perp}^2)
&=& \int \D^2 \Vec k_{\perp} \,
\frac{\Vec P_{h \perp}^2 \Vec k_{\perp}^2 -
 (\Vec k_{\perp} \cdot \Vec P_{h \perp})^2}{\Vec P_{h \perp}^2}
\, f(\Vec k_{\perp}^2, \Vec P_{h \perp}^2, \Vec k_{\perp}
\cdot \Vec P_{h \perp})\,.
\label{i3}
\eq
When deriving the momentum sum rules we use the
analogous identities obtained by
exchanging $\Vec k_{\perp}$ with $\Vec P_{h \perp}$.

\end{document}